\documentclass[fleqn,usenatbib,twocolumn,twocolappendix]{openjournal}

\usepackage{xcolor}
\usepackage{textgreek}
\usepackage[utf8]{inputenc}
\usepackage[english]{babel}
\usepackage{amsmath}
\usepackage{acronym}
\usepackage{url}
\usepackage{xurl}

\usepackage{hyperref}
\hypersetup{
    unicode, 
    colorlinks=true,
    linkcolor=linkcolor,
    citecolor=linkcolor,
    filecolor=linkcolor,
    urlcolor=linkcolor,
}
\usepackage{color,colortbl}
\definecolor{linkcolor}{rgb}{0.0,0.3,0.5}
\usepackage{tensind}
\tensordelimiter{?}
\DeclareGraphicsExtensions{.bmp,.png,.jpg,.pdf}
\usepackage{verbatim}
\usepackage[normalem]{ulem}
\usepackage{orcidlink}
\usepackage{soul}

\DeclareRobustCommand{\VAN}[3]{#2}
\let\VANthebibliography\thebibliography
\def\thebibliography{\DeclareRobustCommand{\VAN}[3]{##3}\VANthebibliography}

\begin{document}
\title[ESCAPE DM TSP overview for astronomers]{Overview of the ESCAPE Dark Matter Test Science Project for Astronomers}

\author{
James Pearson \orcidlink{0000-0001-8555-8561}$^{1}$\thanks{E-mail: james.pearson@open.ac.uk},
Hugh Dickinson \orcidlink{0000-0003-0475-008X}$^{1}$,
Sukanya Sinha \orcidlink{0000-0002-2438-3785}$^{2}$,
and Stephen Serjeant \orcidlink{0000-0002-0517-7943}$^{1}$}

\affiliation{
$^{1}$School of Physical Sciences, The Open University, Milton Keynes, MK7 6AA, UK\\
$^{2}$School of Physics and Astronomy, University of Manchester, Oxford Road, Manchester, M13 9PL, UK
}

\newacro{LSST}{Legacy Survey of Space and Time}
\newacro{OSCARS}{Open Science Clusters' Action for Research \& Society}
\newacro{ESCAPE}{The European Science Cluster of Astronomy \& Particle Physics ESFRI Research Infrastructures}
\newacro{ESFRI}{European Strategy Forum on Research Infrastructures}
\newacro{EOSC}{European Open Science Cloud}
\newacro{OSSR}{Open-source Scientific Software and Service Repository}
\newacro{VRE}{Virtual Research Environment}
\newacro{TSP}{Test Science Project}
\newacro{CERN}{the European Organization for Nuclear Research}
\newacro{LHC}{Large Hadron Collider}
\newacro{Fermi LAT}{Fermi Large Area Telescope}
\newacro{KM3NeT}{Cubic Kilometre Neutrino Telescope}
\newacro{CTA}{Cherenkov Telescope Array}

\begin{abstract}

The search for dark matter has been ongoing for decades within both astrophysics and particle physics. Both fields have employed different approaches and conceived a variety of methods for constraining the properties of dark matter, but have done so in relative isolation of one another. From an astronomer's perspective, it can be challenging to interpret the results of dark matter particle physics experiments and how these results apply to astrophysical scales. Over the past few years, the ESCAPE Dark Matter Test Science Project has been developing tools to aid the particle physics community in constraining dark matter properties; however, ESCAPE itself also aims to foster collaborations between research disciplines. This is especially important in the search for dark matter, as while particle physics is concerned with detecting the particles themselves, all of the evidence for its existence lies solely within astrophysics and cosmology. Here, we present a short review of the progress made by the Dark Matter Test Science Project and their applications to existing experiments, with a view towards how this project can foster complementary with astrophysical observations.

\end{abstract}


\maketitle



\section{Introduction}
\label{sec:introduction}

After the discovery of surprisingly large velocity dispersions for galaxies in the Coma cluster \citep{1933AcHPh...6..110Z}, and following the discovery of discrepancies between spiral galaxy rotation curves and their observed stellar masses \citep{1970ApJ...159..379R}, evidence has continued to mount for a missing mass component in the Universe. This dark matter has long been proposed as the explanation for various physical phenomena, and there have been various possibilities put forward as to its nature \citep{2005PhR...405..279B,2012arXiv1201.3942P}. There are two main types of dark matter: baryonic (primarily non-luminous massive compact halo objects, or MACHOS, such as black holes and brown dwarfs) and non-baryonic (primarily in the form of new particles), with the Universe consisting of a much higher fraction of the latter – the matter density of non-baryonic matter is approximately five times that of all baryonic matter \citep{2020A&A...641A...6P}, and with about 30\% of all baryons accounted for through works by e.g. \cite{2019A&A...624A..48D} the amount of remaining 'baryonic dark matter` implies a ratio of non-baryonic-to-baryonic dark matter of about 7.6. Hence, for the rest of this review we primarily consider only the non-baryonic component. Cold  cold dark matter (CDM) has proven to be the most enduring candidate: such non-relativistic particles have a short free-streaming length than warmer variants, causing them to quickly coalesce under gravity and allowing for the rapid formation of small-scale structures observed in the early Universe \citep{1984Natur.311..517B,2012AnP...524..507F,2014ARA&A..52..291C}.

Dark matter is thought to play a key role in the formation of structure on both galactic and cosmological scales, so to shine a light on its properties requires understanding the dark matter power spectrum and its evolution over cosmic time. Hence, multiple methods are employed to help constrain these properties \citep[][]{2018PhR...761....1B,2022JPhG...49f3001M}. On the largest scales, for example, the power spectrum of Cosmic Microwave Background (CMB) temperature fluctuations is accurately fitted by CDM, with the ratios of peaks providing measures of the relative densities of baryonic and non-baryonic matter in the Universe, among other cosmological parameter constraints. In fitting this power spectrum, results from the Planck Collaboration give an approximately 100$\sigma$ detection of collisionless (i.e. non-baryonic) dark matter \citep{2020A&A...641A...1P,2021PrPNP.11903865A,2025arXiv251005483T}. Meanwhile, x-ray emission from galaxy clusters indicates the presence of large quantities of hot gas that can only be explained should the clusters have a large dark matter component \citep{2002MNRAS.334L..11A}. These dark matter haloes within which galaxies and galaxy clusters reside can also result in cosmic shear that, when modelled, provides constraints on the dark matter density \citep[e.g.][]{2021PhRvL.126n1301T}. Additionally, the alignment of galaxies along our line of sight can produce gravitational lenses, where the foreground galaxy acts as a `lens' (albeit without focusing) around which the background galaxy is distorted and magnified; the modelled dark matter haloes of these lenses contain a central `cuspy' slope \citep[e.g.][]{2021MNRAS.503.2380S,2021A&A...651A..18S} and potential substructures that can be used to constrain the properties of dark matter \citep[e.g.][]{2022MNRAS.510.2464A,2022MNRAS.511.3046H}.

Despite the success of CDM, there still exists a number of astrophysical problems yet to be fully resolved. For example, another source of astrophysical evidence for dark matter comes from the anomalous absorption of the redshifted 21cm line from the early Universe in data from the Experiment to Detect the Global Epoch of Reionization Signature \citep[EDGES,][]{2018Natur.555...67B}. This absorption was found to be over two times greater than their largest predictions, suggestive of significant cooling of primordial hydrogen gas caused by interactions between dark matter and baryons, favouring baryonic dark matter over non-baryonic \citep{2018Natur.555...71B,2018RNAAS...2...37M}. However, this detection has not been subsequently confirmed, such as the non-detection by the Shaped Antenna measurement of the background RAdio Spectrum 3 experiment \citep[SARAS 3,][]{2022NatAs...6..607S}, with other works having sought alternative explanations (see e.g. \cite{2022NatAs...6.1473B} and references therein). Other problems with CDM primarily arise from simulations of CDM haloes. These include the ``core-cusp'' problem \citep{1994ApJ...427L...1F,1994Natur.370..629M}, the ``missing satellites'' problem \citep{1999ApJ...522...82K,1999MNRAS.310.1147M} and subsequent ``too many satellites'' problem \citep{2019MNRAS.487.4409K,2021arXiv210609050K}, and the ``too big to fail'' problem \citep{2011MNRAS.415L..40B,2012MNRAS.422.1203B}. As such, refinements to simulations and further observations from new large-scale surveys like the \textit{Euclid} survey \citep{2011arXiv1110.3193L,2024arXiv240513491E} and the Vera C. Rubin Observatory's Legacy Survey of Space and Time \citep[LSST,][]{2019ApJ...873..111I} are underway to consolidate the properties of dark matter on astrophysical scales.


Particle physics and astroparticle physics offer another, more local approach to understanding dark matter: numerous experiments in particle physics have been seeking to observe dark matter particles themselves, whether by producing dark matter itself, or by detecting it directly or indirectly through its interactions with ordinary matter. For more comprehensive reviews of dark matter candidates, the astrophysical evidence, and the range of particle physics experiments aiming to find them, see the following: \cite{2002LRR.....5....4S,2005PhR...405..279B,2012arXiv1201.3942P,2017arXiv170501987B,2018arXiv180108128B,2018Natur.562...51B,2020JInst..15C6054T,2024arXiv241105062B,2021PrPNP.11903865A,2021NewAR..9301632O,2024arXiv240601705C}, along with the road map for the next decade of US research proposed in the latest Snowmass dark matter complementarity report \citep{10.21468/SciPostPhysCommRep.7}. For detailed reviews of the state of particle physics as a whole, see for example, \cite{2022PTEP.2022h3C01W} and \cite{2024PhRvD.110c0001N}.
Among the leading candidates for dark matter are weakly interacting massive particles (WIMPs): these theoretical CDM particles only interact through gravity and arise from extensions to the Standard Model \citep{2018RPPh...81f6201R}. However, given the above astrophysical problems with CDM and that no clear detections of WIMPs have yet been observed in particle physics experiments, with other theoretical options also being considered \citep{2018Natur.562...51B}. 
These theories include alternative CDM candidates such as axion-like particles \citep{2019arXiv191209123N} and primordial black holes \citep[e.g.][]{2020JCAP...09..022J,2021JPhG...48d3001G,2024NuPhB100316494G}, non-CDM particles such as warm dark matter \citep[WDM; e.g.,][]{2024MNRAS.528.2784D} and self-interacting dark matter \citep[SIDM;][]{2022arXiv220710638A}, as well as modifications to general relativity \citep[e.g.][]{2019JCAP...07..024B}.


\subsection{Synergies in dark matter searches}

Since the first evidence of its existence, the nature of dark matter has remained elusive, and while no dark matter particles have yet been observed, particle physics experiments and astrophysical evidence continue to ever tighten the constraints on its properties.
With increasing search efforts for dark matter underway, efficient coordination and communication between dark matter-related communities is key. As a result, a number of dark matter discussion forums and collaborations have been established that bring together theorists and experimentalists\footnote{An overview of which can be found here: \url{https://www.idmeu.org/dm-related-communities-centers-and-groups/}}. 

The LHC Beyond the Standard Model (BSM) physics Working Group (LHC BSM WG) within the LHC Physics Centre at CERN (LPCC) aims to define guidelines for searches and recommendations for enhancing the reinterpretability of published LHC results. Within this domain, the LHC Dark Matter Working Group (LHC DM WG) focuses on particle physics models for LHC experiments that can highlight the complementarity between collider and non-collider experiments, while the LHC Long-lived Particles Working Group (LHC LLP WG) covers the physics of new long-lived particles and unconventional experimental signatures from dark matter and dark sector scenarios. 
Within the CERN Physics Beyond Colliders (PBC) Study Group \citep{2020JPhG...47a0501B}, the Feebly Interacting Particles Physics Centre (FPC) has been providing a forum for exchanges between the PBC experimental community and theorists, and developing the potential of the PBC experiments for the physics of feebly-interacting particles also by taking into account results from neighbouring fields like dark matter direct detection, astroparticle physics, and cosmology. 

The European Consortium for Astroparticle Theory \citep[EuCAPT;][]{2021arXiv211010074A} aims to coordinate ideas, activities, resources and open environments for the European community of theoretical astroparticle physicists and cosmologists. 
The iDMEu project \citep[the initiative for Dark Matter in Europe and Beyond;][]{2024taup.confE.333C} aims to provide a common platform to facilitate both cross-community dark matter discussions and the collection of resources in an online meta-repository, supported by ECFA (the European Committee for Future Accelerators), NuPECC (the Nuclear Physics European Collaboration Committee) and APPEC (the Astroparticle Physics European Consortium).

\subsection{Dark matter and open science}
\label{subsec:open-science}

With ever-increasing data volumes of current and next-generation facilities, there is also a growing need for coordination and communication between research infrastructures, including across different domains of physics. Additionally, sustainability of these projects is required for scientific reuse, as concerns grow over the reproducibility of results in science: for example, a previous study showed that, across physics and engineering, 70\% of researchers were unable to reproduce others' results, and 50\% were unable to reproduce their own results \citep{2016Natur.533..452B}. As such, platforms are needed to host and publish data, analyses, and software to ensure accountability, accessibility, reinterpretability, and long-term reproducibility in accordance with open science principles \citep{euopenscience}.
For dark matter searches, we highlight the Dark Matter Data Centre \citep[DMDC;][]{2023chep.confE.305B} within the ORIGINS Data Science Laboratory (ODSL) as one such platform, including data sets, workflows, and interactive visualisations, with databases maintained on the Max Planck Computation \& Data Facility GitLab\footnote{\url{https://www.origins-cluster.de/odsl/dark-matter-data-center}}.

One of the European Union's Open Science enablers is the \ac{EOSC}, which was created to provide a multidisciplinary environment for open research in Europe, where researchers can make use of tools and services to store and reuse the data and results of their and others' work according to FAIR (Findability, Accessibility, Interoperability and Reusability) principles.
To support these open and FAIR practices of \ac{EOSC}, the \ac{OSCARS} project has been established to bring together world-class European Research Infrastructures (RIs), connecting scientific communities and supporting collaborations, towards advances in open science -- this review is supported by the OSCARS project. 
These RIs pertain to five Science Clusters: Humanities and Social Sciences; Life Sciences; Environmental Sciences; Photon and Neutron Science; and Astronomy, Nuclear and Particle Physics. 
In particular, the latter is covered by \ac{ESCAPE}, which includes next-generation RI facilities within the astronomy, astroparticle and particle physics communities. These RIs are especially concerned with challenges of data-driven research: \ac{ESCAPE} developed federated storage, data services and infrastructure to accommodate this, including the ESCAPE Data Lake, the Software Catalogue \ac{OSSR}\footnote{\url{https://zenodo.org/communities/escape2020/}}, and the \ac{VRE} analysis platform\footnote{\url{https://github.com/vre-hub}, and \url{https://vre-hub.github.io/}} \citep[see, e.g.,][]{Gazzarrini:2023zax, 10.3897/arphapreprints.e116673}.


With funding from the EOSC-Future project \citep{bird_2021_6390607}, ESCAPE developed the VRE and delivered the Dark Matter Test Science Project\footnote{\url{https://eoscfuture.eu/data/dark-matter/}} \citep[TSP,][]{2021thep.confE..29C}, connecting some of the RIs within ESCAPE that involve searches for dark matter. 
The Dark Matter TSP was established to demonstrate ESCAPE services and open science capabilities, enabling cross-talk between different experiments across astrophysics and  particle physics, and delivering new scientific results in terms of dark matter searches. 

To date, the focus within this TSP has been on particle physics experiments (see Section~\ref{sec:dm-tsp} for more details), with the TSP developing tools for the high energy particle physics and astroparticle physics communities to, for example, visualise the constraints on dark matter particle properties from various experiments. However, it is worth emphasising that all evidence for dark matter's existence is astrophysical in nature, yet there remains a disconnect between astronomers and the particle physics communities. The tools and workflows in the TSP do not as yet provide information interpretable for astronomy, nor are astrophysical constraints on dark matter integrated into these tools. This lack of shared tools and services prevents astronomers from understanding and utilising results and constraints from particle physics, and vice versa. As such, in this review we seek to address these concerns, primarily to help bridge this divide from an astronomy perspective.
In this work, we provide an overview of the Dark Matter TSP, including the tools developed and the research utilising them in particle physics experiments. We do this from the point of view of astronomers, so that members of the astronomical community may understand and provide their own constraints on dark matter, and with the hope that they may use the ESCAPE tools themselves to provide up-to-date research/constraints on the nature of dark matter.

This paper is organised as follows. In Section~\ref{sec:dm-tsp}, we provide an overview of the dark matter TSP, including the various particle physics experiments it supports and common tools available on the VRE. Section~\ref{sec:astrophysical-constraints} presents explanations of example astrophysical observations of dark matter that could be incorporated into the TSP. Software tools available for astrophysical dark matter constraints, and for relating these to particle physics constraints, are presented in Section~\ref{sec:tools}, followed by concluding remarks in Section~\ref{sec:conclusion}.


\section{The Dark Matter Test Science Project}
\label{sec:dm-tsp}

Through making use of ESCAPE tools and services hosted on the EOSC, the Dark Matter TSP seeks to store, distribute, and provide FAIR software and data access for dark matter research in order to highlight synergies between different research communities and allow them to collaborate to produce new results \citep{2021thep.confE..29C}. In particular, the project focuses on experimental data and software from the following direct detection, indirect detection, and particle collider experiments: the ATLAS general-purpose particle detector experiment; the DarkSide direct detection experiment; and the \ac{KM3NeT}, \ac{Fermi LAT}, and \ac{CTA} indirect detection experiments. Theoretical and observational constraints are also to be used, with the aim of combining these data analyses in a coherent way, storing data and software on the ESCAPE Data Lake and OSSR respectively, and providing access to the data analysis pipeline through the ESCAPE VRE (where analyses for the above are already accessible\footnote{\url{https://github.com/vre-hub/science-projects/tree/main}}).
In this section we provide an overview of the above experiments, as well as how the dark matter TSP has contributed to each and to combining their results.

One of the goals of the the Dark Matter TSP for comparing constraints and highlighting the complementarity of different experiments is to host the end-to-end workflows to produce the curves necessary for dark matter summary plots, which display individual experimental constraints from the outputs of each particle physics experiment workflow that can be interpreted in terms of the dark matter candidate properties \citep{bird_2021_6390607}. These plots are commonly used in dark matter experiments and typically show constraints between dark matter candidate mass $m_{DM} {\rm (GeV)}$ and a cross-section $\sigma {\rm(cm^2)}$ describing the probability of a certain interaction during a collision between that candidate and a given particle, such as those sketched for the dark matter annihilation and WIMP-nucleon scattering cross-section in Figures 2 and 3 of \cite{10.21468/SciPostPhysCommRep.7}. This interaction (scattering or annihilation) cross-section can be multiplied by the particle flux to give the interaction rate, i.e. the number of interactions per unit time. Depending on the model, different cross-sections can be represented, such as the spin-independent interaction cross-section $\sigma_{SI}$ or the thermally averaged annihilation cross-section $\langle \sigma v \rangle {\rm(cm^3\ s^{-1})}$. The latter is velocity-weighted, i.e. averaged over all dark matter velocities, and describes the average rate at which annihilation processes occur when multiplied by the dark matter number density \citep{2024arXiv241105062B}.

\subsection{The ATLAS particle detector}

Particle physics experiments seek to observe the presence of new particles directly from high-energy collisions between baryonic matter, and to detect the resulting particles either directly or from their decay products. Under the assumption that dark matter can couple (interact) with Standard Model particles in some way, at the \ac{LHC} at \ac{CERN}, the general-purpose ATLAS experiment aims, among its other physics objectives, at producing invisible particles such as dark matter and infer their existence from their decay into visible matter. 

Each component of the ATLAS detector is specifically designed to record and identify different Standard Model particles. Quarks and gluons coming out of the LHC collisions give rise to a large number of hadrons which can be reconstructed into cone-shaped collider objects, called jets. The only Standard Model particle which does not interact with any of the detector components is the neutrino, since they are colour and charge neutral, and have faint weak interactions. In a collider hadron detector environment, only the plane transverse to the beam axis is of interest, and total transverse momentum of all particles has to be zero after collision, since the initial particles (protons) move along the beam axis. Missing transverse momentum (also termed missing transverse energy, or MET) is the transverse momentum carried away by non-detectable particles and thus `missing' to cancel an observed net momentum in the direction transverse to the collider beam axis. The presence of neutrinos in a collision event causes an imbalance of transverse momentum and contributes to event MET. If, however, a collision only produces invisible particles (such as neutrinos or dark matter) particles, it would also go undetected through the detector. This is a challenge when probing for dark matter models in a collider experiment.

\subsubsection{Dilepton resonance search}

In simplified models of dark matter interactions with SM particles \citep{Abercrombie:2015wmb}, the interactions between dark matter and SM particles are mediated by new massive particles. An example of such a mediator is the massive gauge boson labelled as a $Z^{\prime}$ boson (akin to the existing $Z^0$ boson in the Standard Model). These mediators can also decay back into a pair of SM particles, such as leptons or quarks, leading to a peak or `resonance' in the invariant mass Standard Model continuum for these particles.
One of the searches considered in the Dark Matter TSP is a dilepton resonance search, looking for these new mediator particles -- since no new signal was found over the background in the data collected to date, constraints were set on the fiducial cross-section of the $Z^{\prime}$ particle \citep{2019PhLB..796...68A}, which could be extrapolated to the full data set expected to be collected over the lifetime of the LHC \citep{2018ATLAS}.

The Dark Matter TSP aimed to reinterpret this search using updated models of dark matter mediators, and to do so in a reproducible way using the tools developed by ESCAPE. Summary plots showing constraints on the fiducial cross-section of a new $Z^{\prime}$ particle as a function of mass were produced in the VRE and included in the US `Snowmass process' reports \citep{2022arXiv220312035A,2022arXiv220913128B,2022arXiv220603456B,2022arXiv221001770B}, the community vision for the next decade of particle physics research.

\subsubsection{$t$-channel semi-visible jet search}

Several theories propose that dark matter manifests as part of a complex group of particles in a hidden sector~\citep{Strassler:2006im}, akin to the Standard Model, governed by a new ``dark force'' -- a framework known as dark Quantum Chromodynamics (dark QCD). This force would explain interactions between the dark matter particles themselves, as well as between dark and Standard Model particles. These models target a non-WIMP scenario, giving rise to unusual and unexplored collider event-topologies. One such collider signature is termed as semi-visible jet, where parton evolution includes dark sector emissions, resulting in jets interspersed with dark matter particles. The total momentum of the dark matter is hence correlated with the momentum of the visible states, leading to the direction of MET being aligned close to a jet. 
If dark mesons exist, their evolution and hadronization procedure are currently little constrained. They could decay promptly and result in a very Standard Model QCD-like jet structure~\citep{Park:2017rfb}, even though the original decaying particles are dark sector ones; they could behave as semi-visible jets~\citep{Cohen:2015toa,Cohen:2017pzm}; or they could behave as completely detector-stable hadrons, in which case the final state is just the missing transverse momentum. Depending on the lifetime of the dark mesons, they could appear to ``emerge'' within the detector volume, termed as emerging jets~\citep{Schwaller:2015gea}. 

There have been initial searches for these models at the ATLAS experiment~\citep{ATLAS:2023kao,ATLAS:2023swa,ATLAS:2025kuz,ATLAS:2025bsz}. One of these ATLAS searches has probed the semi-visible jet signature in $t$-channel production mode, details of which can be found in work by~\cite{ATLAS:2023swa}. No new signal was found over the background of known particles for the collisions tested. Assuming a coupling strength of unity between the scalar mediator, a Standard Model quark and a dark quark, mediator mass limits were obtained. Additionally, upper limits on the coupling strengths were also derived. Owing to the broad range of possible collider signatures originating from these models, this search was designed in a generalised manner, using moderate kinematic selections. This enables the search to be reinterpreted in the context of a wider range of dark matter models which might have similar collider final-state signatures. 

Separate to the Dark Matter TSP, the ATLAS Collaboration has since implemented the $t$-channel semi-visible jet search into ESCAPE services with the motive of analysis preservation in mind, and so the search is now fully available through the VRE. This analysis implementation hence prototyped and demonstrated the use of the VRE as a long lasting service existing beyond the scope of the Dark Matter TSP.

\subsection{DarkSide direct detection}

Rather than directly producing dark matter, direct detection experiments aim to observe interactions of pre-existing dark matter particles with baryonic matter. As such interactions, if they exist, are extremely rare, large-scale experiments under carefully controlled conditions are needed to increase the likelihood of a detection. Under the assumption that dark matter consists of WIMP-like particles that can interact with baryonic matter to produce nuclear or electron recoil, the DarkSide experiment utilises a large chamber of liquid argon to search for both scintillation and ionisation that result from the elastic scattering of argon nuclei.
Operating in the underground Gran Sasso National Laboratory (LNGS) in Italy, the latest iteration of the detector, DarkSide-50, contains almost 50kg of liquid argon within its central dual-phase time projection chamber, using ultra-pure argon from underground sources to minimise the abundance of its radioactive isotope \citep{2015PhLB..743..456A,2016PhRvD..93h1101A}.

From the data, the Dark Matter TSP has implemented reanalysis tools for high-mass searches on the ESCAPE VRE, producing DarkSide-50 exclusion curves for the WIMP-nucleon cross-section.
They have also worked on implementing a low-mass analysis, as well as developing their tools to allow for different theoretical models to be inserted by users in order to produce different constraints on dark matter.
Such tools may well be of use in the near future: the DarkSide collaboration is now also building a larger detector called DarkSide-20k, containing tens of tonnes of liquid argon, to further extend the discovery potential of the direct detection program \citep{2023arXiv231203597M,2024CmPhy...7..422T}.

\subsection{Gamma ray \& neutrino indirect detection}

Another method of constraining dark matter is through various indirect detection methods, which infer the presence of dark matter from the visible-matter end products of its decay or annihilation, such as photons, electrons, and neutrinos. The dark matter TSP has focused on two main indirect detection methods, gamma rays and neutrinos, searching for an excess of these secondary particles above the expected background. Such methods are made challenging by complex astrophysical backgrounds that require sophisticated statistical and computational techniques to remove.

\subsubsection{Fermi LAT}
\label{subsec:fermi_lat}

The \textit{Fermi} LAT is a space-based gamma ray detector that has been operating in the MeV to TeV range for more than a decade, scanning the entire sky every $\sim$192 minutes from the low-Earth orbit. One faint source of gamma rays could be from dark matter particle annihilation or decay that, while rare, would be most prevalent in denser dark matter regions, such as galaxy centres, galaxy clusters, and dwarf spheroidal galaxies \citep{2012PDU.....1..194B}. The \textit{Fermi} LAT's high angular and energy resolutions and its comparatively large effective area should allow it to detect such faint emissions \citep{2016PhR...636....1C}, whose energies scale directly with the dark matter particle mass.

One project prior to ESCAPE, MLFermiDwarfs, used real measurements to train machine learning models to predict the gamma ray background over the entire sky \citep{2018JCAP...10..029C,2020JCAP...09..004A}. This aimed to remove foreground mismodelling in the data from Milky Way dwarf spheroidal galaxies, and to provide a more robust framework to derive constraints on the velocity-independent dark matter annihilation cross-section. For the Dark Matter TSP, the results and software tools used to produce them were reproduced and implemented on the VRE\footnote{\url{https://gitlab.in2p3.fr/escape2020/virtual-environment/mlfermilatdwarfs}}, along with data moved to the ESCAPE Data Lake, making them publicly accessible \citep{calore_2021_5592836} and optimised to allow for customisation and quick checks of the viability of user-defined dark matter models. 
A similar process was also done for the analysis of gamma ray flux limits from dark matter capture rates in 13 nearby cold and old brown dwarfs by \cite{2023PhRvD.107d3012B}. Through scattering interactions, these objects are hypothesised to accumulate dark matter particles which can in turn annihilate into lighter mediator particles. If these particles are then long-lived enough to decay into photons once outside the brown dwarf, they can be detected. 
The results showed that the current sensitivity of \textit{Fermi} is not high enough to enable bounds to be set on the dark matter-nucleon elastic scattering cross-section, requiring a factor of 9 improvement in the upper limits on the gamma-ray flux to be able to achieve bounds of $\sim$$10^{-36} \text{cm}^2$ for dark matter masses below 10~GeV \citep{2024PhRvD.109l9904B}.
The software was again made open source \citep{bhattacharjee_2024_11519115} and fully available through the VRE\footnote{\url{https://gitlab.in2p3.fr/escape2020/virtual-environment/brown-dwarfs-gamma}} to allow others to extend the work to other astrophysical objects.

\subsubsection{KM3NeT + CTA}

The KM3NeT neutrino telescope is a detector located deep underwater in the Mediterranean Sea, featuring the low-energy detector ORCA (Oscillation Research with Cosmics in the Abyss) and high-energy detector ARCA (Astroparticle Research with Cosmics in Abyss) which can be combined depending on the particle mass and decay channel of interest. Consisting of digital optical modules (DOMs; photomultiplier tubes arranged in glass spheres) positioned along vertical flexible strings, ORCA will feature 115 100m-long strings at a depth of $\sim$2500m, while ARCA will eventually have two sets of 115 $\sim$700m-long strings each, collectively covering a $1{\rm km}^3$ volume anchored at a depth of 3500m. Submerged in sea water that acts as a shield against atmospheric muons, the DOMs measure Cherenkov radiation produced by secondary particles from any interactions of neutrinos with the water. Such neutrinos may be produced from the decay or annihilation of dark matter from astrophysical sources, and may be accompanied by other signals: KM3NeT/ARCA is hence used for multimessenger astronomy, offering an accurate means of source detection in dense source regions (neutrinos only weakly interact so are negligibly deflected by matter on their path through space) that can be combined with electromagnetic or gravitational wave experiments that provide information about the energy spectrum and time window, respectively.

Here, the Dark Matter TSP has focused on the telescope's instrument response function (IRF), with regard to estimating rates expected for detection events and the background. For example, this can be used to show the relationship between energy resolution, effective area/volume and angular resolution of the detector, and can help circumvent the need for complex simulations. The tool developed for this was added to the VRE\footnote{\url{https://gitlab.in2p3.fr/escape2020/virtual-environment/irf-from-km3net}}, as it may also be applicable to other experiments and the VRE enables remote execution of what is a computationally expensive task. As an example, similarities exist between gamma ray and neutrino astronomy which allowed the tool to be deployed \citep{unbehaun_2023_8298464} as part of a combination of data from KM3NeT with the high-energy ground-based gamma ray experiment CTA, and used to distinguish between different emission scenarios of gamma ray sources in the Milky Way \citep{2024taup.confE.227S,2024EPJC...84..112U}.

\subsection{Common tools on the VRE}
\label{subsec:vre-tools}

One of the main benefits of the EOSC and VRE is the ability to share common `off the shelf' tools and algorithms that can be of benefit to various research projects and across research communities. For example, with increasing numbers of large-scale facilities and projects, the storage of vast amounts of data is becoming an ever-growing issue across communities, especially for particle physics and astronomy. Recognising this challenge, the Dark Matter TSP have developed and implemented on the VRE a prototype of a machine learning-based data compression tool called \textsc{Baler} \citep{2023arXiv230502283B,alexander_ekman_2024_10723669} as an example of a reusable tool solving a common problem. \textsc{Baler} can be used to test the feasibility of compressing different types of scientific data using autoencoders, including training and testing a model, saving the resulting model and compressed data, and decompressing the model at a later date and plotting the performance.

\textsc{Baler} joins a number of other software packages available through the ESCAPE OSSR and VRE that are of use to astronomers\footnote{\url{https://zenodo.org/communities/escape2020/}}. For example, the platform includes \textsc{Aladin Lite} \citep{2022ASPC..532....7B}, a browser-based astronomical HiPS visualiser; \textsc{Gammapy} \citep{2023A&A...678A.157D}, a Python toolbox for gamma-ray astronomy; and a series of Jupyter Notebook tutorials on using astronomical databases and Virtual Observatory tools \citep{marchand_2025_14720244}.

In summary, the Dark Matter TSP has developed multiple open source codes and workflows that run in the VRE, which are still being used for dark matter research, including producing up-to-date summary plots as new results come in from the various dark matter direct detection, indirect detection, and particle collider experiments.


\section{Translating to Astrophysical Constraints}
\label{sec:astrophysical-constraints}

So what does all of this mean for astronomers? Having discussed some of the particle physics experiments and the constraints they place on dark matter properties, we now shift focus to examples of how these could relate to observable astrophysical constraints. While here we focus primarily on extragalactic tests for these examples, there exist a range of other probes on galactic and cosmological scales that could also be related to these constraints and incorporated into the VRE: for reviews, see e.g. \cite{2018PhR...761....1B} and \cite{2022JPhG...49f3001M}, including Figures 1 and 3 in the former that depict dark matter candidates in astronomically relevant parameter spaces.

\subsection{Gravitational Lensing}
\label{subsec:grav-lensing}

As mentioned in Section~\ref{sec:introduction}, the multiple distorted and magnified lensed images of a background galaxy (`source') around a foreground gravitational lens provide an observable way of measuring the distribution of matter and dark matter within the lens. For extended sources, whose images are both magnified and distorted into large arcs or Einstein rings, any substructure within the lens dark matter halo in turn produces perturbations in the lensed images that can be observed \citep{2016ApJ...823...37H}. Meanwhile for unresolved sources like quasars, whose multiple lensed images are purely magnified in the form of an Einstein cross, lens substructure impacts the ratios of the fluxes of these images \citep{2023MNRAS.524.6159K}. Work has also been done to combine both effects to increase the sensitivity to substructure \citep{2024MNRAS.533.1687G}. 

Different dark matter models are expected to produce differing numbers and properties of subhaloes: for example, compared to CDM, warm dark matter (WDM) particles have higher thermal velocities at early times and hence a larger free-streaming length that prevents small-scale structure from forming \citep[e.g.][]{2001ApJ...556...93B,2016ApJ...823...37H,2023OJAp....6E..39A}, 
reducing the number of low-mass $(<10^{9}\ M_{\odot})$ subhaloes with a cut-off in the halo mass function that varies with the inverse of the particle mass \citep[on the order of ${10^8\ M_{\odot}}$ for keV-scale particle masses;][]{2022MNRAS.511.3046H}.
Careful modelling of these subhaloes, using complex mass models within tools like \textsc{lenstronomy} \citep{2018PDU....22..189B,2021JOSS....6.3283B} or \textsc{PyAutoLens} \citep{2021JOSS....6.2825N}, to reproduce the observed perturbations and flux ratios can therefore be used to place limits on dark matter free-streaming length and the particle masses of CDM \citep[e.g. the 4.1~keV lower bound from][]{2022MNRAS.511.3046H} and WDM \citep[e.g. the 2.0~keV lower bound from][]{2017JCAP...05..037B}, with WDM models expected to produce fewer small-scale perturbations than CDM \citep{2023arXiv230611781V}. To date, only a few lenses have been studied in this way for low-mass haloes \citep[e.g.][]{2010MNRAS.408.1969V,2012Natur.481..341V,2016ApJ...823...37H,2020MNRAS.492.3047H}, however it is estimated that around 50-100 lenses with accurately measured substructure may be enough to set sufficient limits on WDM mass and potentially rule out CDM if no lower-mass haloes are detected \citep{2016MNRAS.460..363L,2019MNRAS.487.5721G,2019BAAS...51c.153S}.

Another proposed family of models for dark matter is self-interacting dark matter (SIDM), for which such particles can have non-gravitational interactions that exchange energy and momentum, rather than remaining collisionless \citep{2000PhRvL..84.3760S,2022arXiv220710638A}. Whether these interactions consist of elastic or inelastic scattering depends on the specific model, as does whether the interaction cross-section is constant or velocity dependent \citep{2012MNRAS.423.3740V,2013MNRAS.430...81R,2018MNRAS.476L..20R,2021MNRAS.501.4610R}. Such interactions between dark matter particles in the central halo and accreting subhaloes allow for ram-pressure stripping of the latter, with an efficiency that scales with the self-interacting cross-section \citep{2023arXiv230611781V}. This stripping suppresses the numbers of lower-mass subhaloes compared to CDM: for example, see the peak velocity function of model subhaloes in Figure 6 of \cite{2020ApJ...896..112N}, which shows the abundance of surviving SIDM subhaloes becoming increasingly suppressed compared to CDM simulations at lower peak velocities. As such, observations and simulations of gravitational lensing can be used to test these models, constraining the self-interaction cross-section at low velocities.

Within the high-density regions of haloes and subhaloes of SIDM, the self-interaction cross-sections also affect their central mass density profiles, with some experiencing core-collapse from the transfer of heat in these interactions and forming cuspy cores \citep{2014PhRvL.113b1302K,2018MNRAS.479..359S,2019MNRAS.484.4563D,2022MNRAS.513.4845Z}. As such, observations of gravitational lenses can explore the potential diversity of these core distributions that would arise from SIDM \citep{2019MNRAS.488.3646R}. 
For example, \cite{2021MNRAS.507.2432G} showed that flux ratios from quadruple-image strongly lensed quasars enable probing of self-interactions at velocities below ${\rm 30\ km\ s^{-1}}$, with 50 such objects having the potential to rule out CDM depending on the measured interaction cross-section amplitude at such low velocities. Such experiments need to use quasar emission at wavelengths in which the projected source-plane sizes are of the order milliarcseconds or larger, to minimise microlensing effects by stars within the foreground lens \citep[e.g.][]{2024MNRAS.530.2960N}. 
Meanwhile, simulations from \cite{2025PhRvD.111f3001Z} showed that the core-collapse of subhaloes exhibits unique observable features in lensing, with a SIDM cross-section of ${\rm \geq 200\ cm^2\ g^{-1}}$ typically required for a significant fraction of subhalos to core-collapse.
However, gravitational lens modelling suffers from a number of degeneracies that can, for example, lead to over- or under-estimation of the number of low-mass subhaloes, a review of which is presented in \cite{2023arXiv230611781V}. Nevertheless, in the coming decade these will no doubt be accounted for through the continued development of simulations paired with high-resolution imaging of orders of magnitude more lenses following wide-field surveys like LSST and \textit{Euclid}.

\subsection{Dwarf Spheroidal Galaxies}
\label{subsec:dwarf-spheroidal-galaxies}

As mentioned in Section~\ref{subsec:fermi_lat}, dwarf spheroidal galaxies (dSph) are considered to be promising targets for indirect dark matter detection using ground- and space-based $\gamma$-ray telescopes \citep{2018RPPh...81e6901S,2024PhRvD.110f3034A}. 
Measurements of stellar velocity dispersions in dSphs reveal very high measured mass-to-light ratios and therefore indicate high densities of dark matter are present. Moreover, dSphs contain relatively little gas and relatively few stars, which results in a low flux of background astrophysical $\gamma$-rays \citep{2022JCAP...11..055A}. This makes distinguishing a faint dark-matter signal significantly less challenging than it would be in the direction of the Milky Way centre, for example.

High energy $\gamma$-rays are an expected product from the annihilation or decay of WIMP-like dark-matter particles. The $\gamma$-ray signal is expected to be characterised by continuum emission resulting from hadronization of decay products (including, e.g., $W^{\pm}$ bosons, quark-antiquark pairs or electron-positron pairs) and subsequent pion decay, or from direct decays yielding one or two $\gamma$-ray photons. While the latter process would produce a ``smoking gun'' line signal in the observed $\gamma$-ray spectrum, its flux is expected to be fainter than the continuum from hadronization by a factor $1/\alpha^{2}$ where $\alpha$ is the electromagnetic fine structure constant. This is because WIMP-like particles cannot couple directly to photons and so Feynman diagram for the annihilation to photons must include a virtual charged particle loop.

For WIMP annihilation, the expression for the expected flux ${\mathrm{d} \Phi_\gamma}/{\mathrm{d}E_{\gamma}}$ of $\gamma$-rays produced with energy $E_{\gamma}$ and arriving from within a solid angle $\Delta\Omega$ can be written as the product of two terms \citep[e.g.][]{1998APh.....9..137B,2022JCAP...11..055A},

\begin{equation}
\frac{\mathrm{d} \Phi_\gamma}{\mathrm{d}E_{\gamma}}(E_{\gamma},\Delta\Omega)=\frac{1}{4 \pi}\underbrace{\sigma_\gamma \frac{\left\langle\sigma_{\text {ann}} v\right\rangle}{2 m_{\text {DM}}^2} \sum_f \frac{\mathrm{dN}_{\gamma}^f}{\mathrm{\text{d}E_{\gamma}}}\mathrm{B}_f}_{\text {Particle Physics}}\,\times\, \underbrace{J(\Delta\Omega)}_{\text{Astrophysics}}\label{eq:gr_flux_generic}
\end{equation}

The first term in Equation~\eqref{eq:gr_flux_generic} encapsulates parameters and quantities describing the particle physics of dark matter annihilation, including the velocity-averaged annihilation cross section $\left\langle\sigma_{\text {ann }} v\right\rangle$, the WIMP mass $m_{\text {DM}}$, the expected spectrum ${\mathrm{dN}_{\gamma}^f}/{\mathrm{\text{d}E_{\gamma}}}$ of $\gamma$-rays with energy $E_{\gamma}$ produced by a specific annihilation channel $f$, and the branching ratio for that channel $B_{f}$, where the sum covers all possible annihilation channels.

The second term in Equation~\eqref{eq:gr_flux_generic} is often referred to as the \mbox{``J-factor''} and describes an integral within a solid angle $\Delta\Omega$ along the observer's line of sight of the square of the dark matter density distribution. Typically, dSphs appear point-like at the spatial resolution of $\gamma$-ray telescopes, and so $\Delta\Omega$ is taken to be equal to the telescope's effective beam size.

By assuming that the annihilation cross-section is independent of the relative velocity between the annihilating dark-matter particles (as is the case for s-wave annihilation\footnote{In quantum mechanics, a scattering process can be solved through a partial wave expansion, which decomposes the process into components and treats it as the scattering of constituent waves with defined angular momentum quantum numbers, $l$, such as s- ($l=0$), p- ($l=1$), and d-waves ($l=2$). Hence, only the first few need taking into account for low-energy scattering processes.}) and that density of the dSph's dark-matter halo has a radially symmetric profile $\rho_\text{DM}$, then the J-factor can be written as an integral along the line-of-sight (los) coordinate $s$ and over the solid angle $\Delta\Omega$.
\begin{equation}
    J(\Delta\Omega) = \int_{\Delta\Omega} \text{d}\Omega \int_\text{los}\rho_\text{DM}^{2}(s)\,\text{d}s
    \label{eq:j_fac_swave}
\end{equation}

More generally, the annihilation cross-section may depend on the dark matter particles' relative velocities (e.g. for p-wave annihilation or for SIDM) and many dark matter models, including WDM, CDM and SIDM, predict that the dSph halo may have complex substructure.

In such cases, the J-factor depends on the detailed distribution of dark matter density along the line of sight. Unresolved dark matter substructures with higher density than the bulk dSph halo can significantly enhance (potentially by an order of magnitude) the J-factor when averaged over $\Delta\Omega$ since it must be the case that $\langle\rho^{2}\rangle \geq \langle\rho\rangle^{2}$.

Measurements of the locations and radial velocities of stars in dSphs can be used to constrain the form and normalisation of $\rho_\text{DM}$ in Equation~\eqref{eq:j_fac_swave} \citep[e.g.][]{2020ApJ...904...45H}. However, making these measurements can be very challenging because dSphs are often very faint and contain very few stars. Consequently, estimates of the J-factor derived from astrophysical measurements are subject to significant systematic uncertainties, particularly when the stars belonging to a dSph are difficult to separate from foreground and background interlopers.

Ultimately, neither of the terms in Equation~\eqref{eq:gr_flux_generic} is well constrained and unknown halo substructure parameters further complicate matters. To infer the physical properties of dark matter using indirect detection methods, one must typically make assumptions about the astrophysical parameters encapsulated in the J-factor (for example, whether the dark matter is clumpy or not), albeit that some of those assumptions can be constrained using observational data. Using these observationally constrained assumptions, one may compare the predicted $\gamma$-ray spectrum for various specific dark matter particle models and compare with that which is observed to determine the most likely model, given the available data. 


\section{Tools for Dark Matter Constraints}
\label{sec:tools}

In addition to the online platforms and repositories mentioned in Section~\ref{subsec:open-science} and the software discussed in Section~\ref{subsec:vre-tools}, various tools have been developed over the years to facilitate the above astrophysical dark matter constraints and relate these to particle physics constraints, a few examples of which are presented here. With sufficient interest, tools like these could be incorporated into the VRE to facilitate streamlined astrophysical dark matter searches.

\subsection{\textsc{pyHalo}}
\label{subsec:pyhalo}

Strong lens modelling can be used to constrain WDM and CDM halo properties such as the mass-concentration relation \citep{2012MNRAS.424..684S,2020MNRAS.492L..12G} when combined with software that can simulate full mass distributions with substructure, for instance the \textsc{pyHalo} Python package\footnote{\url{https://github.com/dangilman/pyHalo/}} \citep{2020MNRAS.491.6077G}. This code can also be used to constrain the self-interacting cross sections of SIDM, which is implemented by defining mass bins for subhaloes and field haloes and specifying the fraction of core-collapsed haloes in each bin (which have different density profiles to non-collapsed haloes). This is then passed to a lens modelling package, primarily \mbox{\textsc{lenstronomy}}\footnote{\url{https://github.com/lenstronomy/lenstronomy}} \citep{2018PDU....22..189B,2021JOSS....6.3283B}, to perform ray tracing to compute the effective gravitational distortion produced by all of the haloes and produce the resulting lensed images. Building on the work discussed in Section~\ref{subsec:grav-lensing}, \cite{2023PhRvD.107j3008G} applied this method to study quadruply imaged quasars, from which their models disfavoured cross sections exceeding ${\rm 100\ cm^2\ g^{-1}}$ at relative velocities below ${\rm 30\ km\ s^{-1}}$, for which most haloes undergo core-collapse: they obtained the mass-binned core-collapse fractions based on a characteristic timescale of halo evolution that partially depends on the thermally-averaged cross section, halo mass, and redshift \citep{2023ApJ...946...47Y,2023ApJ...949...67Y}.

\subsection{\textsc{GAMBIT}}
\label{subsec:gambit}

We have seen that there are many experiments and astrophysical observables for constraining dark matter properties, and the combination of multiple approaches will be necessary to provide concrete evidence for a given theory. The \textsc{GAMBIT}\footnote{\url{https://gambitbsm.org/}} \citep[Global And Modular BSM Inference Tool;][]{2017EPJC...77..784A} software is a global fitting code for theories going Beyond the Standard Model (BSM), used to simultaneously analyse data from many sources. The code performs statistical global fits of such BSM models by combining theoretical predictions of observables with targeted searches across a wide range of experimental data from particle physics and astrophysics, computing observables and likelihoods alongside various statistical interpretations of results such as goodness-of-fit p-values and Bayes factors for model comparisons. 

Along with a backend for dynamical interfacing with external tools used to compute physical quantities, \textsc{GAMBIT} also consists of several modules (or `Bits') designed to provide native simulations for collider and astrophysics experiments, two of which we mention here: \textsc{DarkBit} \citep{2017EPJC...77..831B} and \textsc{CosmoBit} \citep{2021JCAP...02..022R}. 
The former module is designed for computing dark matter observables and likelihoods for multiple direct and indirect detection experiments, including interfacing with external packages to calculate relic densities \citep{2020JPhCS1342a2059C}, and has been used to explore a range of dark matter candidates: see \cite{2025EPJWC.31911002B} for a review. 
Regarding the latter module, many BSM scenarios also have cosmological implications that are missed when fitting purely to particle physics experiments, and likewise many cosmological theories beyond $\Lambda$CDM can potentially produce new detectable signals in such experiments.
As such, \textsc{CosmoBit} has been developed to combine cosmological and particle physics data simultaneously to better constrain theories, computing cosmological observables and likelihoods for Type Ia supernovae, large-scale structure, Big Bang Nucleosynthesis and the cosmic microwave background. It offers a flexible framework for beyond-$\Lambda$CDM theories to be tested, such as modifications to inflation, particle properties, and the effective number of relativistic degrees of freedom, and the code has already allowed for the first global analysis of the parameter space of axion-like particles, whose decay into photons would affect various astrophysical and cosmological observables \citep{2022JCAP...12..027B}.

\subsection{\textsc{AxionLimits} and the \textsc{Dark Matter Limit Plotter}}
\label{subsec:dm-limit-plotter}

Of course, once a dark matter model has been tested, the resulting constraints need comparing to pre-existing limits, typically done visually by way of dark matter summary plots as mentioned in Section~\ref{sec:dm-tsp}. With these graphs being so commonplace in particle physics, \textsc{AxionLimits}\footnote{\url{https://github.com/cajohare/AxionLimits}} hosts files and Python notebooks for creating summary plots for axions, axion-like particles, dark photons, and other ultralight bosons \citep{AxionLimits}. Additionally, the \textsc{Dark Matter Limit Plotter}\footnote{\url{https://supercdms.slac.stanford.edu/science-results/dark-matter-limit-plotter}} developed by the Super Cryogenic Dark Matter Search (SuperCDMS) at the SLAC National Accelerator Laboratory presents an interactive dashboard to generate customisable figures. 
Users can also apply for their results to be uploaded and included in the plotter for others to use, and as such it already contains many pre-existing limits from various experiments to which any new results can be compared.


\section{Conclusions}
\label{sec:conclusion}

There are currently several particle physics experiments and astrophysical observations that are being used to test a vast array of theoretical dark matter models. Acknowledging that many of the ESCAPE Research Infrastructures form part of this search, and with an ever-growing need to compare and combine the results from these approaches, the ESCAPE Dark Matter TSP was established to provide a platform to facilitate this in an open and FAIR way as part of EOSC. The project's analyses and tools are already aiding scientific communities in producing new constraints from particle detectors and direct and indirect detection experiments within particle physics, as well as ensuring reproducibility and providing a testing ground for future experiments' software and computing infrastructure.

In this review, we have provided an overview of the ESCAPE Dark Matter TSP, as well as some observational astronomical dark matter searches, and corresponding tools frequently used in the analyses of dark matter constraints. The review has been aimed primarily at astronomers rather than particle physicists, in order to provide an introduction to the dark matter searches currently incorporated in EOSC, which may share complementarity with observational astronomical dark matter searches. Additionally, to date the TSP has been limited to particle and astroparticle physics experiments but now seeks to expand its reach to incorporate astronomical dark matter searches into its services, for example to produce astronomically relevant and interpretable summary plots like those mentioned at the start of Section~\ref{sec:astrophysical-constraints}.

In providing this review, we encourage astronomers to make use of the ESCAPE services within the Dark Matter TSP to facilitate their own dark matter searches, enabling an open, collaborative pathway towards combining complementary constraints from astronomy and particle physics to maximise our understanding of dark matter over the coming decade.

\section*{Acknowledgments}

The authors gratefully thank Caterina Doglioni, Kay Graf, Tetiana Hryn'ova, Valerio Ippolito and Francesca Calore for their helpful and insightful comments and suggestions. JP, HD and SSe are all supported by the ACME, ELSA, and OSCARS projects. ``ACME: Astrophysics Centre for Multimessenger studies in Europe'', ``ELSA: Euclid Legacy Science Advanced analysis tools'', and ``OSCARS: Open Science Clusters' Action for Research and Society'' are funded by the European Union under grant agreement no. 101131928, 101135203, and 101129751, respectively; ELSA is also funded by Innovate UK grant 10093177. SSi is supported by European Research Council grant REALDARK (Grant Agreement no.~101002463). Views and opinions expressed are however those of the authors only and do not necessarily reflect those of the European Union.


\section*{Data Availability}

No new data were generated or analysed in support of this research.


\bibliographystyle{mnras}

\bibliography{bibliography}







\end{document}